\documentclass[aps,prb,preprint,groupedaddress]{revtex4-1}
\usepackage{xspace}
\usepackage{amsmath}
\usepackage{epstopdf}
\usepackage{graphicx}

\begin{document}


\title{Non-conservative current-driven dynamics: beyond the nanoscale}


\author{Brian Cunningham}
\email{b.cunningham@qub.ac.uk}
\author{Tchavdar N. Todorov}
\author{Daniel Dundas}
\affiliation{Atomistic Simulation Centre, School of Mathematics and Physics, Queen's University Belfast, Belfast BT7 1NN, U.K.}



\begin{abstract}
Long metallic nanowires combine crucial factors for non-conservative current-driven atomic motion. These systems have degenerate vibrational frequencies, clustered about a Kohn anomaly in the dispersion relation, that can couple under current to form non-equilibrium modes of motion growing exponentially in time. Such motion is made possible by non-conservative current-induced forces on atoms, and we refer to it generically as the waterwheel effect. Here the connection between the waterwheel effect and the stimulated directional emission of phonons propagating along the electron flow is discussed in an intuitive manner. Non-adiabatic molecular dynamics show that waterwheel modes self-regulate by reducing the current and by populating modes nearby in frequency, leading to a dynamical steady state in which non-conservative forces are counter-balanced by the electronic friction. The waterwheel effect can be described by an appropriate effective non-equilibrium dynamical response matrix. We show that the current-induced parts of this matrix in metallic systems are long-ranged, especially at low bias. This non-locality is essential for the characterisation of non-conservative atomic dynamics under current beyond the nanoscale.
\end{abstract}


\maketitle


\section{Introduction}

The development of electronic devices at the nanoscale is a challenging avenue of research with the aim of improving their efficiency and performance. This requires an understanding of the mechanisms for energy transfer from current carriers into atomic motion. Large current densities can generate significant additional forces on atomic nuclei \cite{brandbyge_origin_2003,di_ventra_current-induced_2002,todorov_current-induced_2001}, resulting in a class of phenomena known as electromigration: atomic rearrangements and mass transport driven by current flow \cite{landauer_driving_1974,sorbello_theory_1997}. Recent work has drawn attention to another aspect of these forces, anticipated in a visionary argument by Sorbello \cite{sorbello_theory_1997}: unlike equilibrium interatomic forces, they are non-conservative (NC), enabling the current to do work on atoms around closed paths \cite{dundas_current-driven_2009,stamenova_current-driven_2005,lu_blowing_2010,bode_scattering_2011}.

This mechanism for energy conversion from current into atomic motion -- which we refer to as the waterwheel effect -- differs from Joule heating \cite{horsfield_power_2004,galperin_molecular_2007} in two key respects. First, the growth in atomic kinetic energy is exponential. Second, it is not stochastic: the energy transferred in the waterwheel effect is stored in directional motion -- specifically -- as generalised angular momentum \cite{todorov_nonconservative_2011}. In the early work above it seemed that the waterwheel effect might require rather specialised conditions. The effect operates fundamentally through the coupling of pairs of normal modes to form generalised rotors driven by the current. This requires modes that are close in frequency and are,  furthermore, strongly coupled by the NC current-induced forces.

A class of systems where these requirements are met are long, low-dimensional metallic wires \cite{cunningham_nonconservative_2014}. They have a dense frequency spectrum providing the desired degeneracies. In addition, frequency renormalisation by the current (which in general can ruin the degeneracies) is small in these quasi-ballistic systems. Finally, electrons couple strongly to extended phonon modes with the wavevectors needed for momentum conservation under backscattering \cite{agrait_onset_2002}. Simulations under current indeed show NC dynamics in long atomic wires on a grand scale \cite{cunningham_nonconservative_2014}.

This study revisits the waterwheel effect in long wires, and reports on two further aspects of this problem. The first is the physical interpretation of the effect. Originally the effect was demonstrated for a system with just two degrees of freedom -- the corner atom in an atomic wire with a bend \cite{dundas_current-driven_2009}. Under the right conditions, current drives the atom around an expanding orbit in analogy with a real waterwheel, enabling an intuitive picture of how NC forces work. We will see that an intuitive analogy at the other end of the spectrum, i.e., in extended systems, is also possible: it is how strong winds generate forward-travelling ripples on a lake, or the uncompensated stimulated emission of directional phonons \cite{todorov_nonconservative_2011,lu_current-induced_2015}.

However, this process hinges on momentum conservation, and for waves this information requires a sufficiently long-ranged physical property. For atomic motion under current this property is the non-equilibrium dynamical response matrix, whose anti-symmetric part (induced by the current) describes the NC forces \cite{lu_blowing_2010,dundas_ignition_2012}. The second advance reported here is the quantitative analysis of this property of long metallic nanoconductors. We show that in long nanowires this anti-symmetric part becomes very long-ranged. This non-locality is essential for the characterisation of NC dynamics under current beyond the nanoscale.

\section{Methods}

\begin{figure}
\includegraphics[width=12cm,trim=0cm 0cm 0cm 0cm]{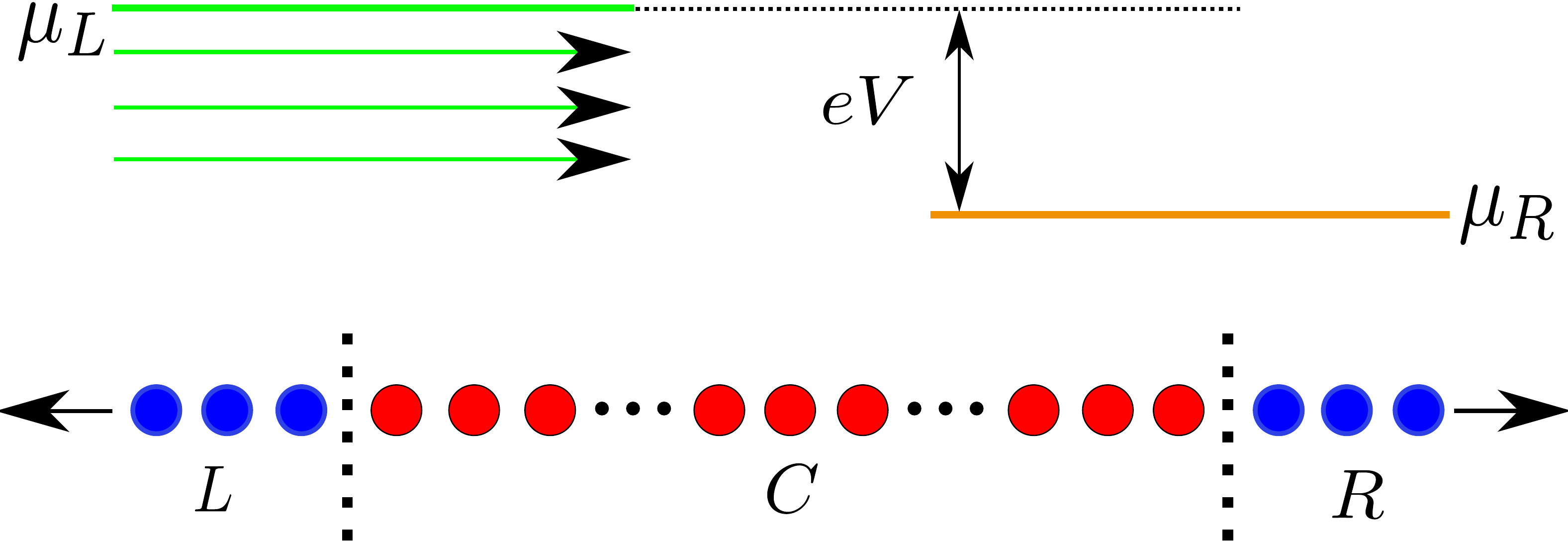}
\caption{Long one-dimensional (1D) atomic chain. Region $C$ (red) is composed of 200 mobile atoms sandwiched between 50-atom long rigid regions; on the outside, $C$ is then connected to electrodes $L$ and $R$ (blue). Electrons are injected into the electrodes by sources held at different electrochemical potentials, $\mu_L$ and $\mu_R$. For $\mu_L>\mu_R$, net particle current flows from left to right.}
\label{fig:system}
\end{figure}

The system investigated is illustrated in Fig.~\ref{fig:system}: a central region $C$, whose middle section containing 200 atoms will be treated dynamically, and two electrodes $L$ and $R$ to generate current flow.

We are interested in the behaviour of the ions (nuclei and core electrons) in these systems under current, and in particular, the gain in atomic kinetic energy due to work done by the current. The force we are considering is the {\sl mean} force exerted by electrons on ions. It is determined by the rate of change of the expectation value of the ionic momentum. Using the Ehrenfest approximation, the force on an atomic degree of freedom $n$ is given by
\begin{equation}
\boldsymbol{F}_n=-{\rm Tr}\left\{\hat{\rho}_e \nabla_n\hat{H}_e(\vec{R})\right\},
\label{force}
\end{equation}
where $\hat{\rho}_e$ is the one-electron density matrix and
$\hat{H}_e(\vec{R})$ is the one-electron Hamiltonian
as a parametric function of the atomic positions $\vec{R}$.

We will employ two different approaches to determine $\hat{\rho}_e$: an adiabatic steady-state approach,
where $\hat{\rho}_e = \hat{\rho}_e(V,\vec{R})$ is a function of the bias $V$ and the geometry $\vec{R}$,
and a non-adiabatic dynamical approach (within the mixed quantum-classical Ehrenfest method), where
$\hat{\rho}_e = \hat{\rho}_e(t)$ is obtained from an open-boundary quantum Liouville equation
\cite{mceniry_dynamical_2007}. These approaches will be discussed in more detail later.

Electrons are described within a spin-degenerate single-orbital orthogonal nearest-neighbour tight-binding model with parameters fitted to the elastic properties of bulk gold \cite{sutton_simple_2001}.
The nearest-neighbour Hamiltonian matrix elements have the form
\begin{equation}
H_{e,mn}=-\frac{\epsilon c}{2}\left(\frac{a}{R_{mn}}\right)^q,
\end{equation}
where $R_{mn}$ is interatomic distance,
$\epsilon=0.007868~{\rm eV}$, $a=4.08~{\rm eV}$, $c=139.07$, and $q=4$.
In addition, the model includes a repulsive pair potential of the form
\begin{equation}\label{eq:pair}
P_{mn}=\epsilon \left(\frac{a}{R_{mn}}\right)^p,
\end{equation}
with $p=11$. The onsite elements of the Hamiltonian are set to zero,
and the electron band-filling is $\nu = 0.36361$.
Non-interacting electrons are considered throughout.
As in \cite{cunningham_nonconservative_2014} we compress the chain to
a lattice spacing of $R=2.373~{\textup{\AA}}$, to suppress a Peierls distortion
and resultant band gap that form after geometry relaxation.

\subsection{Landauer steady state}

In the adiabatic steady-state method for the electronic structure,
we employ the Landauer picture of conduction. Here the electrodes are infinite,
and electrons populate sets of stationary Lippmann-Schwinger scattering states,
arriving from either side. The respective populations functions $f_{L}(E)$ and
$f_{R}(E)$ correspond to the electrochemical potentials of the left and right
source reservoirs. The steady-state electron density matrix is then given by
\cite{todorov_tight-binding_2002}
\begin{equation}
\hat{\rho}_e=\int_{-\infty}^{+\infty}\left\{f_L(E)\hat{D}_L(E)+f_R(E)\hat{D}_R(E)\right\}{\rm d}E,
\end{equation}
where $\hat{D}_i(E)$, $i=L,R$ are the density of state operators -- subsuming spin degeneracy -- for
the two sets of scattering states. We work at zero electronic temperature where
the occupations are step functions. The density of states operators are generated
by Green's function techniques.

We can now calculate the forces on ions about a chosen reference geometry $\vec{R}_0$.
Under small displacements ${\rm d}\vec{R}$;
\begin{equation}
F_n(\vec{R})\approx F_n(\vec{R}_0)-\sum_m K_{nm}{\rm d}R_{m},
\end{equation}
where $K_{nm}=-\partial F_{n}(\vec{R}_0)/\partial R_{m}$ are the elements
of the dynamical response matrix.

The dynamical response matrix determines the vibrational frequencies and
corresponding collective modes of motion of the ions. We ignore velocity-dependent
forces in the present steady-state description (although they will be present in the
non-adiabatic dynamical simulations later).  These forces can be included perturbatively \cite{lu_blowing_2010,todorov_current-induced_2014} and tend to dampen the atomic motion,
and introduce a contribution arising from the Berry phase \cite{lu_blowing_2010}.
Force noise is also excluded here.

The vector containing the atomic displacements can then be expressed as
\begin{equation}\label{eq:eqom_L}
{\rm d}\vec{R}(t)=\sum_j\vec{p}_j\left\{A_je^{{\rm i}\omega_j t}+B_je^{-{\rm i}\omega_jt}+\widetilde{f}_j\right\},
\end{equation}
where $\{\vec{p}_j\}$ are the eigenvectors of the dynamical response matrix,
with frequencies $\{\omega_j\}$, and the static contribution $\{\widetilde{f}_j\}$
is determined by any residual forces present in the chosen reference geometry.
$\{A_j\}$ and $\{B_j\}$ are set by initial conditions.

The dynamical response matrix can be separated into an equilibrium and a current-induced part:
\begin{equation}
\boldsymbol{K}=\boldsymbol{K}_{\rm eq}+\Delta\boldsymbol{K}.
\end{equation}
The current-induced part, in turn, is separated into a symmetric and anti-symmetric part \cite{dundas_ignition_2012}, $\Delta\boldsymbol{K}=\boldsymbol{S}+\boldsymbol{A}$, with
\begin{align}
S_{nm}=\frac{\Delta K_{nm}+\Delta K_{mn}}{2}~\\
A_{nm}=\frac{\Delta K_{nm}-\Delta K_{mn}}{2}.
\end{align}
The anti-symmetric part, present only under bias, is a generalisation of the curl of the force on an ion \cite{dundas_ignition_2012,cunningham_nonconservative_2014}. The resultant non-hermiticity of the dynamical response matrix under current in general generates complex frequencies. The complex modes come in complex conjugate pairs. Via Eq.~(\ref{eq:eqom_L}) these modes give rise to solutions that grow or decay exponentially in time. These are the waterwheel mode pairs investigated in \cite{dundas_current-driven_2009,todorov_nonconservative_2011,dundas_ignition_2012,cunningham_nonconservative_2014}.

Within this steady-state approach, current is determined from the bias and the reference geometry, and is not allowed to respond to the subsequent motion of the ions. This approach is accurate for small atomic displacements and large atomic mass (suppressing the velocity-dependent forces relative to the non-conservative forces, and also the work rate due to inelastic scattering \cite{montgomery_power_2002}), and in systems where fluctuations in the current due to deviations from ideal steady-state behaviour are not too large \cite{cunningham_nonconservative_2014}.

\subsection{Dynamical simulations}

To simulate departures from the above ideal conditions, we use the
non-equilibrium non-adiabatic molecular dynamics method of \cite{mceniry_dynamical_2007}.
Now the leads are finite and embedded in external electron baths supplying carriers.
The leads in the present simulations will be 250 atoms long, with open-boundary parameters $\Gamma=0.5~{\rm eV}$ and $\Delta=0.0005~{\rm eV}$ \cite{mceniry_dynamical_2007}.
The electron density matrix then evolves according to the open-boundary equation of motion
with the source and sink terms present \cite{mceniry_dynamical_2007}, in the presence of the atomic motion. Atoms obey Newtonian equations, with forces found from the
time-evolving density matrix via Eq.~(\ref{force}).
This form of electron-ion dynamics is known as Ehrenfest dynamics. It captures all
forces -- equilibrium and non-equilibrium -- with the exception of the force noise associated
with spontaneous phonon emission and Joule heating \cite{todorov_current-induced_2014}.

By contrast with the Landauer method above, the dynamical simulations accommodate
departures from steady-state conditions and allow the current to respond to changes in the
vibrational amplitudes.


To compare the two methods we will further perform a short-time Fourier
transform on the ion trajectories $\vec{R}(t)$ from the dynamical simulations
and examine the evolution of the energy distribution across the phonon band.
The Fourier transform uses a Blackman window, effectively suppressing data
outside a particular time interval, while ensuring the data remains continuous.
The frequency spectrum of the total ionic kinetic energy (for all $N$ ions) is then
\begin{equation}\label{eq:energy_mode}
E(\omega) \propto \omega^2\sum_{n=1}^{N}\left|A_n(\omega)\right|^2,
\end{equation}
where $\{A_n(\omega)\}$ is a Fourier component of $\{R_{n}(t)\}$.
The window is then moved along in time.

\section{Results and discussion}
\subsection{Landauer steady-state calculations}

First we analyse the equilibrium vibrational modes determined from the dynamical response matrix for a long metallic wire. The mode analysis is then performed under bias, where complex frequencies are present. Long-range interactions in the dynamical response matrix are investigated.
\subsubsection{Equilibrium mode analysis}
\begin{figure}
\includegraphics[width=0.55\textwidth,clip=true,trim=0.0cm 0cm 0cm 0cm,angle=90]{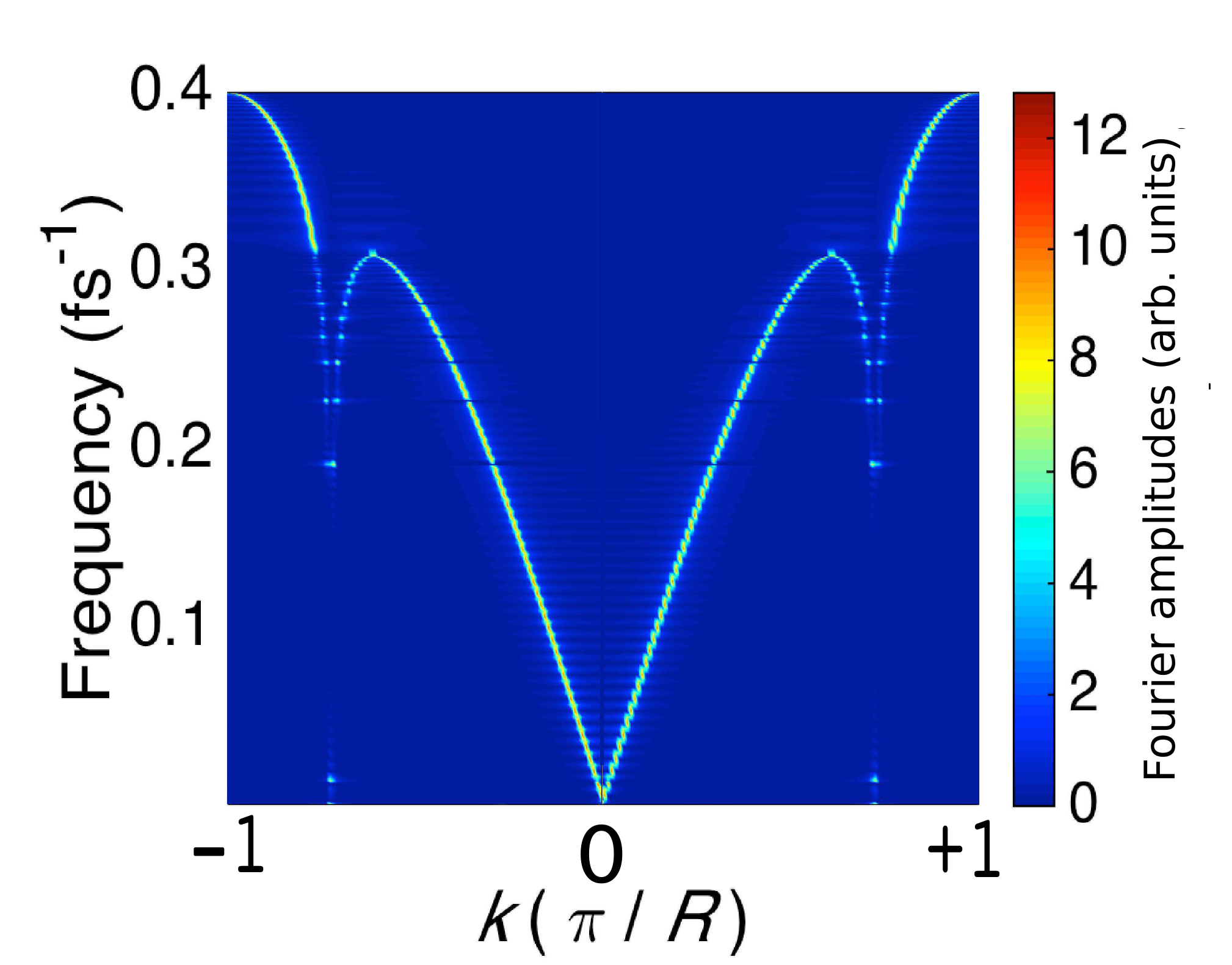}
\caption{The equilibrium frequencies of each vibrational mode (vertical axis).
For each eigenfrequency, the moduli of the $k$-space components (horizontal axis)
of the corresponding eigenvector (normalised to unity) are represented by colour.
The system consists of 200 atoms (relaxed) of mass 10 a.m.u.
sandwiched between two semi-infinite perfect leads.}
\label{image:1dmode_wavevec}
\end{figure}

We analyse the equilibrium eigenfrequencies and eigenmodes for longitudinal phonons in
a chain with 200 mobile atoms of mass $M=10$ a.m.u.. The $200$ eigenvalues $\{k_j\}$
of the $200\times 200$ dynamical response matrix give rise to
$2\times 200$ eigenfrequencies: positive and negative square roots of the eigenvalues divided by mass, $\omega_j = \pm \sqrt{k_j/M}$. Each eigenvector of the dynamical response matrix is of length $200$.
Its elements give the relative amplitudes of the atoms in the given mode.
These real-space eigenmodes are normalised to unity and Fourier transformed into momentum space.  Figure~\ref{image:1dmode_wavevec} presents the mode frequencies (vertical axis), together with
the modulus (colour) of the Fourier components ($k$, horizontal axis)
of the corresponding eigenvector.

Notice the dip around $k=\pm0.73\pi/R$. It arises due to the long range behaviour of the dynamical response matrix. To check this, we have determined the dispersion relation by truncating the dynamical response matrix beyond a chosen cut-off range. The dip appears when the range includes at least third or fourth neighbours. The overall shape of the curve in Fig.~\ref{image:1dmode_wavevec} is also sensitive to the truncation range, with the shape in the figure emerging at about 20 lattice spacings.
The dip in Fig.~\ref{image:1dmode_wavevec} is similar qualitatively to experiment \cite{renker_observation_1973}, and occurs at a wavevector of about $\pm 0.73\pi/R$. This is twice the Fermi wavevector $\kappa_F = \nu\pi/R$. We conclude that this dip is the result of a Kohn anomaly \cite{kohn_image_1959}.

\subsubsection{Mode analysis under bias}

\begin{figure}
\includegraphics[width=0.60\textwidth,clip=true,trim=0.0cm 0cm 0cm 0cm]{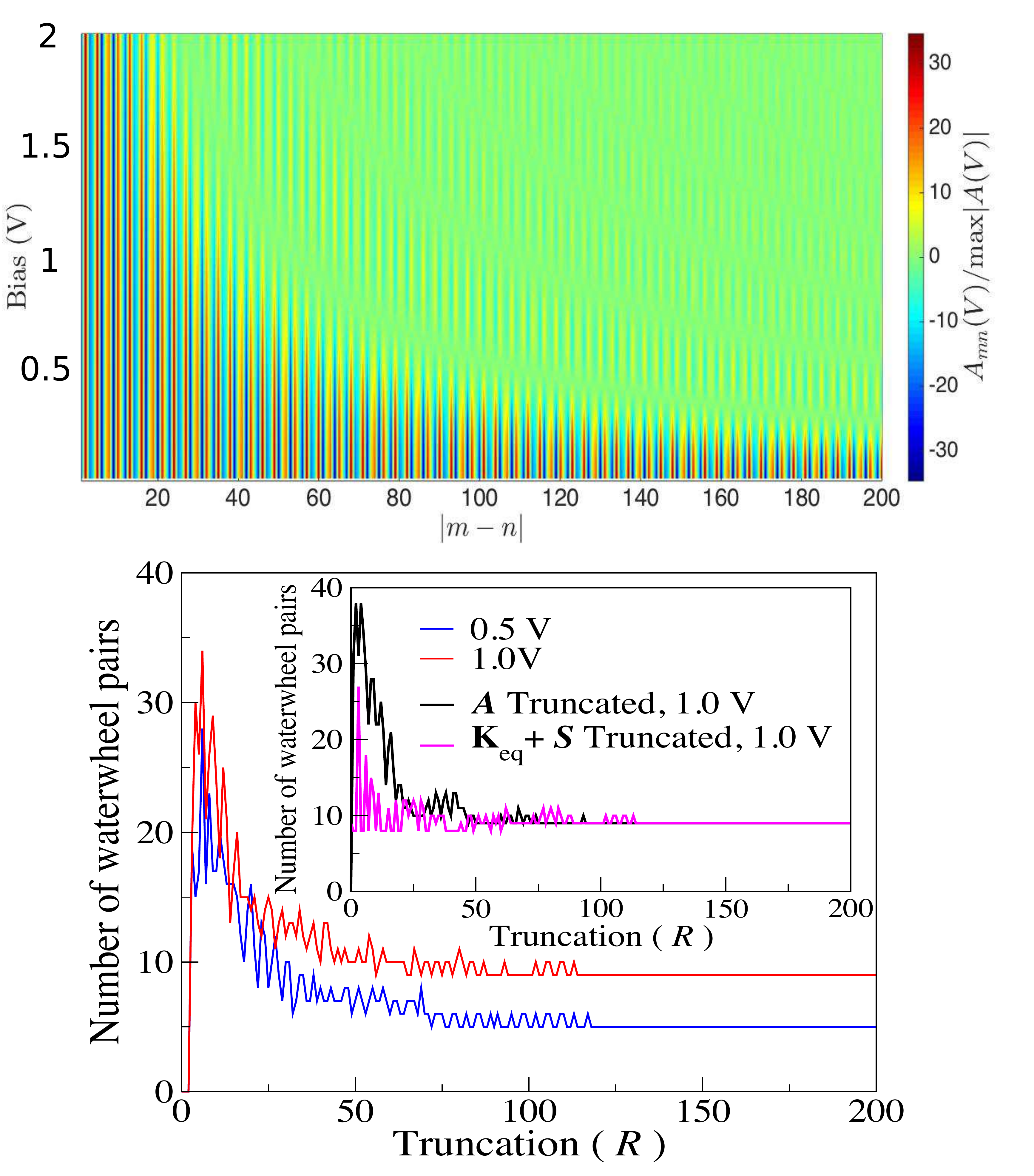}
\caption{Upper panel: for each bias (vertical axis) the elements of the anti-symmetric part of the dynamical response matrix, relative to the maximum element (for the given bias), are presented as a function of site separation. Lower panel: number of waterwheel pairs with imaginary part above $10\%$ of the maximum, as a function of the truncation of the dynamical response matrix. The main panel truncates the whole matrix, $\boldsymbol{K}$, whereas the inset truncates either just the anti-symmetric part (black), or both the equilibrium and non-equilibrium symmetric parts (pink). The system is an infinite perfect wire with 200 mobile atoms.}
\label{fig:range}
\end{figure}

The key difference between the equilibrium and non-equilibrium dynamical
response matrices is the anti-symmetric part of the latter. For an infinite
perfect chain it can be evaluated analytically:
\begin{equation}
A_{mn}=\frac{8q^2\beta}{\pi R^2(4(m-n)^2-1)}\begin{array}{l}\left[\cos{\phi_L}\sin{(2(m-n)\phi_L)}\right.\\\left.-2(m-n)\sin{\phi_L}\cos{(2(m-n)\phi_L)}-f(\phi_R)\right],\end{array}
\end{equation}
where $\beta$ is the hopping integral, $\phi_{L(R)}={\rm cos}^{-1}(\mu_{L(R)}/2\beta)$,
and $f(\phi_R)$ denotes the whole preceding expression in the square brackets with $\phi_L$
replaced by $\phi_R$. The upper panel in Fig.~\ref{fig:range} shows the relative values of the anti-symmetric contribution as a function of site separation and bias.
We see that $A_{mn}$ is oscillatory and
long-ranged and, at small bias, becomes infinitely long-ranged.
The lower panel in Fig.~\ref{fig:range} shows the number of waterwheel pairs
formed under bias for different truncations of the dynamical response matrix.
We see (main panel) that the data deviates from the plateau by a chosen fractional amount,
for longer truncations at low bias; consistent with the upper panel.
The inset then shows that the sensitivity to truncation is set by the non-equilibrium,
anti-symmetric part of the dynamical response matrix.

\begin{figure}
\includegraphics[width=0.6\textwidth,clip=true,trim=0cm 0cm -2cm 0cm]{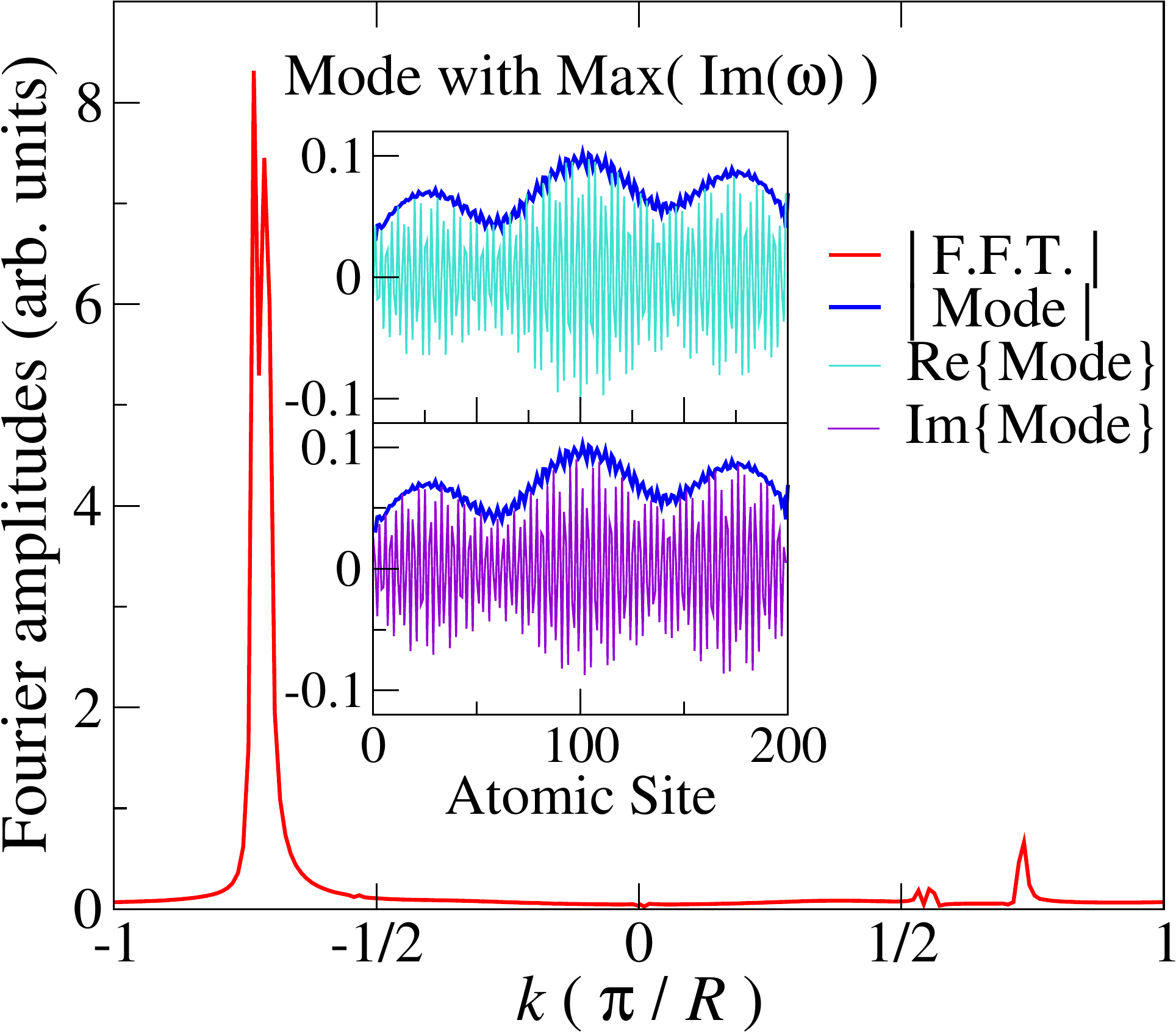}
\caption{Main panel: Fourier components (moduli) of the eigenmode with the largest (negative)
imaginary part to its frequency under 0.5~V. The system consists of 200 mobile atoms,
relaxed under zero bias. Inset: real-space components (moduli) of the eigenmode (blue),
and of their real (turquoise) and imaginary (violet) parts.}
\label{image:1dmode_wavevec_bias}
\end{figure}

We now turn to chains relaxed at zero bias as the reference geometry
(which remain close to the perfect chain). Figure~\ref{image:1dmode_wavevec_bias} presents the eigenvector
of the dynamical response matrix, under bias of $0.5$~V, for the waterwheel mode
with the largest negative imaginary part to its eigenfrequency,
$\omega= 0.237 - 0.083~{\rm i}$~${\rm fs}^{-1}$,
for a wire with 200 mobile atoms (mass 10 a.m.u.).
The inset shows
the real and imaginary parts of the eigenvector (now complex),
together with its modulus;
the main panel displays the moduli of the Fourier
components of the mode in $k$-space. (The Fourier picture is
qualitatively similar for other waterwheel modes.)
First, what motion does this mode describe?
Let $\omega = \omega' + {\rm i}~\omega''$ be the frequency
(with $\omega'$ and $\omega''$ real) and $v_n$ the mode
component at site $n$. The corresponding physical displacement is
of the form
\begin{equation}
R_n (t) = C v_n e^{{\rm i}\omega t} + C^{*} v_{n}^{*} e^{-{\rm i}\omega^{*} t},
\end{equation}
with $C$ an amplitude.
The Fourier spectrum shows that the mode is dominated by a negative $k$-component
(close in magnitude to the value $2\kappa_F$ required for momentum
conservation in electron-phonon interactions, where the Kohn anomaly occurs).
Since $\omega'>0$ and $\omega'' < 0$, we
obtain a right-travelling wave, that grows in time.
The small contribution in the Fourier spectrum
at the corresponding positive $k \sim 2\kappa_F$ describes
phonon waves travelling the other way, that are
attenuated by the electron particle current
(flowing to the right).

These observations can be summarised by the physical picture in the Introduction:
NC forces in long metallic systems generate modes of motion, in which
the current-carrying electrons close to the Fermi level emit a directed shower
of forward-travelling phonons, in analogy with how a breeze whips up waves on a lake.

In addition to the features at $\pm 2\kappa_F$, Fig.~\ref{image:1dmode_wavevec_bias}
shows a weak background of other Fourier components. By mixing in these
other wavevectors, the mode redirects some of the energy gained from
the electron ``wind'' to phonon momenta that can no longer interact directly with the Fermi
electrons, and be reabsorbed. This enables the mode energy to grow in time.

\begin{figure}
\hspace{-0.2cm}
\includegraphics[width=0.9\textwidth,clip=true,trim=1cm 0cm 0cm 0cm]{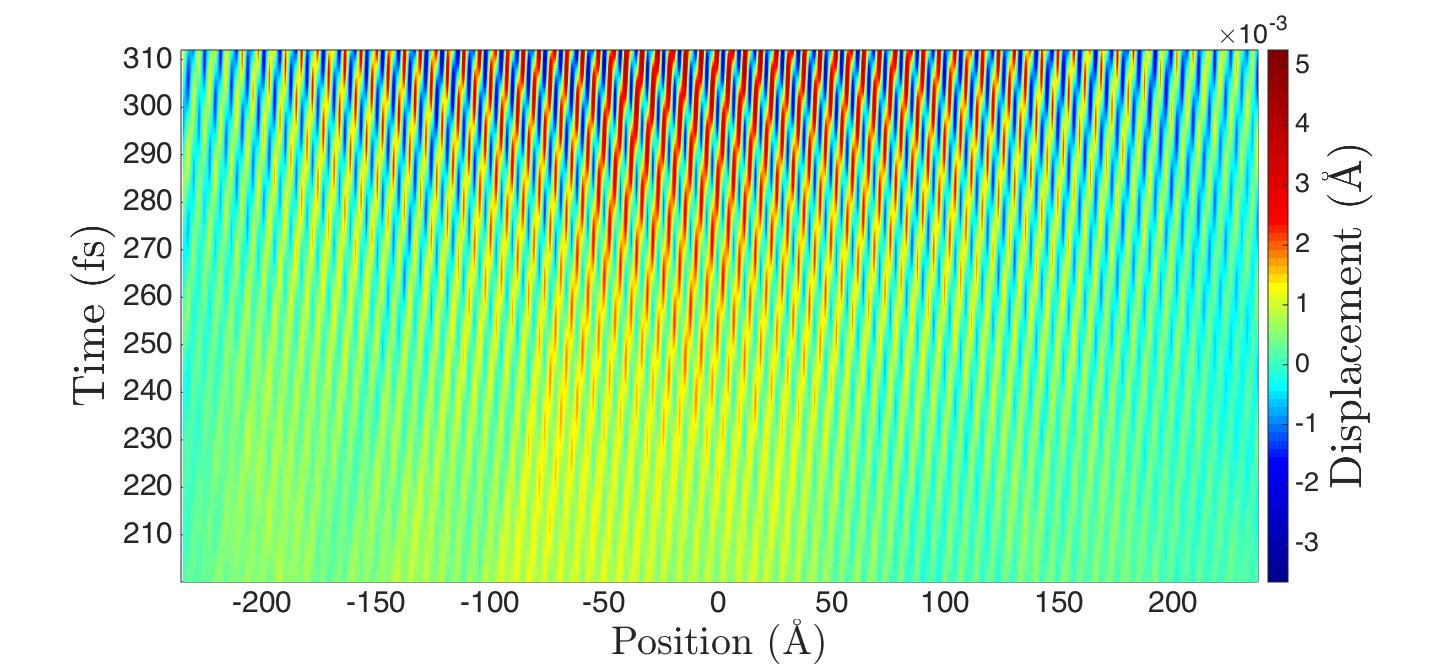}
\caption{Longitudinal atomic displacements in a wire with 200 dynamical
atoms as a function of position (horizontal axis) and time (vertical axis),
for an applied bias of 0.2~V, starting from the zero-bias relaxed geometry.
The simulation employs the small-amplitude adiabatic steady-state
description of Eq.~(\ref{eq:eqom_L}). The atomic mass is 10~a.m.u..}
\label{image:pcol_travel}
\end{figure}

We can use Eq.~(\ref{eq:eqom_L}) to simulate the atomic motion.
We calculate the forces under bias for the zero-bias relaxed geometry,
and use them, with zero initial displacements and velocities,
to set the coefficients $\{A_j\}$ and $\{B_j\}$.
We do not include the friction forces here,
but we cut the imaginary parts of the frequencies
by a factor of 5, to stretch out the growth of the
amplitudes in time.
Figure~\ref{image:pcol_travel} shows the displacements
of the ions as a function of position and time, for a bias of 0.2~V.
The right-travelling phonons
generated by the current are evident.
The velocity of the peaks and troughs is
about $2.5\times10^4~{\rm ms}^{-1}$,
which is approximately equal to $\omega/2\kappa_F$,
with $\omega\sim 0.25~{\rm fs}^{-1}$. This representative
frequency is close to: the Einstein frequency for this system;
the typical real part of the (closely-clustered) waterwheel modes;
the frequency where the dispersion relation (Fig.~\ref{image:1dmode_wavevec})
starts to flatten out; and as we will see,
the dominant frequency observed in the dynamical simulations (below).

\subsection{Non-adiabatic non-equilibrium dynamical simulations}

The dynamical simulations under bias are performed for atoms initially at rest
in the zero bias relaxed geometry. The bias is then ramped up at the start.
The NC forces cause the ionic kinetic energies to increase rapidly. Unlike
the steady-state analysis above however, the current now responds, and is
suppressed by the atomic motion. The system eventually
settles at a dynamical steady state, where the
velocity-dependent friction forces balance out, on average,
the driving NC forces. This interpretation is supported by the fact that balancing these forces leads to analytical predictions that agree with the non-adiabatic simulations \cite{cunningham_nonconservative_2014}. A further, independent verification of this balance is given below. In the simulation below the applied bias is 0.5~V; in the long-time dynamical steady state the current settles at a value corresponding to an effective reduced bias of about 0.2~V (the bias used in the adiabatic visualisation in Fig.~\ref{image:pcol_travel}).

\begin{figure}
\includegraphics[width=12cm,clip=true,trim=0cm 0cm 0cm 0cm]{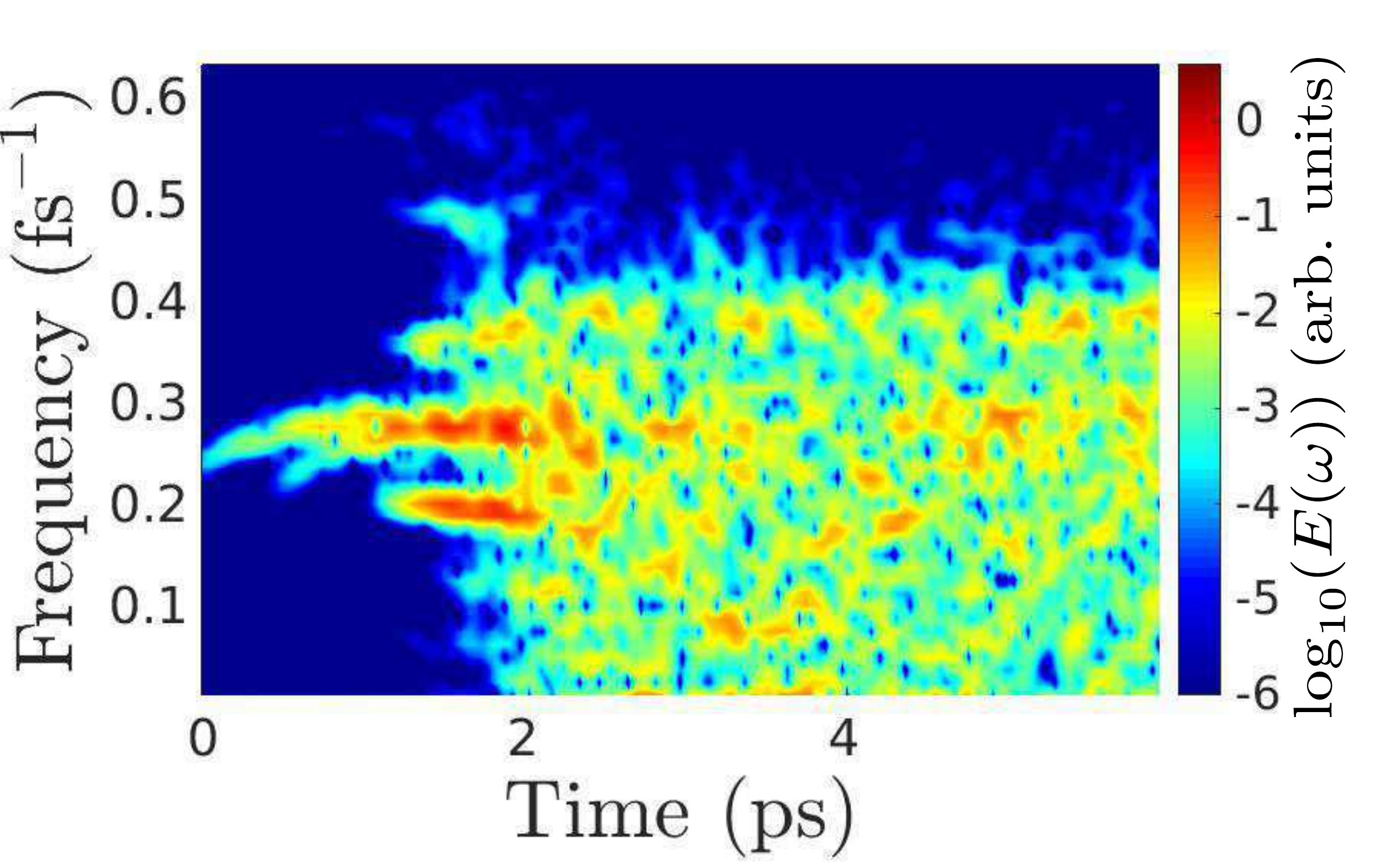}
\caption{Fourier-decomposed total ionic kinetic energy (colour)
across the phonon band, determined using Eq.~(\ref{eq:energy_mode}), during the non-adiabatic dynamical simulation under a bias of 0.5~V, for 200 mobile atoms of mass 10~a.m.u..
A time window of 0.5~ps was applied in increments of
10~fs in the Fourier transform.}
\label{fig:pcol_dyn}
\end{figure}

Figure~\ref{fig:pcol_dyn} presents the temporal Fourier composition of the ionic
kinetic energy, as in Eq.~(\ref{eq:energy_mode}), for the first 6 ps of the simulation.
Initially, the growing
energy is stored in a narrow frequency range,
clustered around the representative frequency above.
As the amplitudes increase, phonon-phonon scattering
eventually redistributes the energy across the phonon band,
resulting in approximate energy
equipartioning among available
frequencies \cite{cunningham_nonconservative_2014}. 

The dynamical simulations can be used to give a measure of the efficiency \cite{PhysRevLett.111.060802} with which the NC forces convert electrical energy into atomic motion. From Ref.~\cite{cunningham_nonconservative_2014} we can estimate the average imaginary part, $\Phi$, of mode frequencies, at a given current. Let $E_0$ be the total atomic kinetic energy in the dynamical steady state (beyond about 2 ps in Fig. \ref{fig:pcol_dyn}). The rate of work by the NC forces is then $W_{\rm NC} = \sim 2E_0\,2\Phi$ (a factor of 2 to give total, as opposed to just kinetic, vibrational energy, and another to convert amplitude to intensity). For the simulation in Fig. \ref{fig:pcol_dyn}, the current in the dynamical steady state is $I \sim 14.2\mu {\rm A}$ and $E_0 \sim 17.5~{\rm eV}$, giving $\Phi \sim 10^{-3}~{\rm fs}^{-1}$ and $W_{\rm NC} \sim 0.07~{\rm eV}{\rm fs}^{-1}$. This can be compared against the power, $W = IV$, due to the transfer of electrons between reservoirs, $W \sim 0.04 ~{\rm eV}{\rm fs}^{-1}$. Thus, for the present systems, these two quantities are comparable. More detailed investigations of this comparison -- including its system-dependence -- is clearly an important direction for further work.

We can also use the above estimates to independently verify the balance between friction and NC forces in the dynamical steady state. For estimation purposes, we use the analytical result for the friction coefficient -- and corresponding energy relaxation time $\tau_{\rm frict}$ -- for atomic Einstein oscillators \cite{mceniry_dynamical_2007}: $1/\tau_{\rm frict} = (4\hbar/M\pi)(H'/H)^2$, for a nearest-neighbour tight-binding chain, where $H$ and $H'$ are the hopping integral and its derivative with distance. For the present parameters, and for the above steady-state kinetic energy, the power lost to friction is $W_{\rm frict} = 2E_0 / \tau_{\rm frict} \sim 0.08~{\rm eV}{\rm fs}^{-1}$, in agreement with the (independent) estimate of $W_{\rm NC}$ above.

\section{Conclusions}

Long low-dimensional metallic systems are a promising testbed for
NC current-driven atomic dynamics. We have highlighted two aspects of
these effects here: the physical interpretation of NC motion as ``ripples''
driven by the electron ``wind'', and the long-ranged character
of the non-equilibrium parts of the dynamical response matrix,
responsible for NC dynamics. The inclusion of Joule heating
(suppressed in Ehrenfest dynamics) and its interplay with the NC forces
is an attractive direction for further work, as is the current-driven
dynamical behaviour in the presence of the Peierls instability
that occurs under compression-free conditions. We hope that this work will motivate further
research into some of these questions.



\begin{acknowledgements}
We are grateful to Jan van Ruitenbeek, Mads Brandbyge, Per Hedeg{\aa}rd,
Jing-Tao L{\" u} and Lorenzo Stella for helpful discussions. We thank the Engineering and Physical Sciences Research Council for support,
under Grant EP/I00713X/1. This work used the ARCHER UK National
Supercomputing Service (http://www.archer.ac.uk).
\end{acknowledgements}

\bibliography{beyond_arxiv}

\end{document}